\documentclass[12pt,showpacs,aps,prd]{revtex4}
\usepackage{graphicx}
\input epsf

\begin{document}
\title{$B^{\ast}B^{\ast}\rho$ vertex from QCD sum rules}
\author{Chun-Yu Cui, Yong-Lu Liu and Ming-Qiu Huang}
\affiliation{Department of Physics, National University of Defense
Technology, Hunan 410073, China}
\begin{abstract}
The form factors and the coupling constants in the $B^{\ast}B^{\ast}\rho$ vertex are evaluated in the framework of three-point QCD sum rules. The correlation functions responsible for the form factors are evaluated considering contributions of both $B^{\ast}$ and $\rho$ mesons as off-shell states. The obtained numerical results for the coupling constants are in agreement with light-cone QCD sum rules calculations.
\end{abstract}
\maketitle

\section{Introduction}
In the last years, both experimental and theoretical studies on heavy mesons have received considerable attention. With growing datasets collected by the collaborations such as CDF, DO, CLEO, and the forthcoming SuperB and LHC, investigations of the spectroscopy and decay of heavy flavor become more exciting~\cite{Klempt:2007cp,Swanson:2006st,Eichten:2007qx,Nora}. Therefore, the reliable determination of various characteristics, such as form factors and coupling constants are needed.

Suppression of heavy quarkonium has been considered for a long time as one of the most striking signatures for the quark-gluon plasma (QGP) formation~\cite{Matsui}. It results from Debye screening of color force in the QGP. In this picture, due to larger
abundances of color charges screening the interaction between the c and $\bar{c}$ quarks, the attractive $c\bar{c}$ potential responsible for the
$J/\psi$ binding gets screened as the temperature of the medium increases in the QGP. (For a detailed review see Refs.~\cite{Nora,Rapp} and references therein.) Extensive experimental efforts have been devoted to study this phenomenon at the Super Proton Synchrotron (SPS) at CERN and Relativistic Heavy Ion Collider (RHIC) at the Brookhaven National Laboratory. As the case of charmonium states in the QGP which are sensitive to the color screening effect, the study of bottomonium suppression in high energy heavy ion collisions can be used as a signature for the QGP as well~\cite{Vogt}. In contrast to charmonium states, bottomonium states should be a cleaner probe of QGP due to its low cross section where the competing effects, which either reduce the yield~\cite{Blaizot} or enhance it~\cite{Andronic}, are negligible. We expect the effects of the QGP on the absorption of $\Upsilon$ in ultra-relativistic heavy ion collisions at the LHC, which can provide an answer to this open question. On the other hand, other more conventional mechanisms based on $J/\psi$($\Upsilon$) absorption by comoving hadrons have also been proposed as a possible explanation~\cite{Cassing,Armesto,linko2}. In this way, one needs to understand the effects of $\Upsilon$ absorption in hadronic matter. It is known that $\pi$ and $\rho$ are the dominant hadrons in ultra-relativistic heavy ion collisions. Since it is still difficult to study strong interaction phenomena at non-perturbative regime using the QCD, the study of quarkonium absorption is generally performed in the framework of effective Lagrangian with meson exchange~\cite{linko2}. In the process, $\Upsilon$ and $\rho$ produce the final states $B^{\ast}$ and $B^{\ast}$ by exchanging a $B^{\ast}$ meson. The calculation of $\Upsilon$ absorption cross sections needs information of the $B^{\ast}B^{\ast}\rho$ interactions. To describe the strong interactions of the negative-parity heavy mesons with $\rho$ meson, we employ an effective Lagrangian, which is constructed based on the chiral symmetry and described by the following~\cite{LZH1}:
\begin{eqnarray}
{\cal L}&=& ig_{BB\rho}Tr\left [\left(
B^\dagger\buildrel\leftrightarrow\over\partial_\mu
B\right)P^\mu\right]-2f_{B^*B\rho}\varepsilon^{\mu\nu\alpha\beta}Tr\left[\left(B^\dag\buildrel\leftrightarrow\over{\partial_\mu}B^{*}_\nu-
B^{*\dag}_\nu\buildrel\leftrightarrow\over{\partial_\mu}
B\right)\partial_\alpha
P_\beta\right]\nonumber \\
&+&ig_{B^*B^*\rho}Tr\left [\left(\bar
B^{*\dagger}_\mu\buildrel\leftrightarrow\over\partial_\nu
B^{*\mu}\right)P^\nu\right]+4if_{B^*B^*\rho}m_{B^*}Tr\left [\left(
B^{*\dagger}_{\mu}B^*_{\nu}\right)\left(\partial^{\mu}P^{\nu}-\partial^{\nu}P^{\mu}\right)\right],
\label{1}
\end{eqnarray}
where $B$ and $B^*$ represent isospin doublets, $P$ is the isospin triplet of the $\rho$ meson. The $B^{\ast}B^{\ast}\rho$
interactions are characterized by two indepentent coupling constants $g_{B^{\ast}B^{\ast}\rho}$ and $f_{B^{\ast}B^{\ast}\rho}$. It is necessary to know values of coupling constants with some precision. The choice of a lower or higher value may change the final cross section to some extent.

Theoretically, the knowledge of the heavy-heavy-light mesons coupling constants in hadronic vertices are very important in estimating strength of hadron interactions when hadronic degrees of freedom are used. They are fundamental objects of low energy QCD. They may also play an important role in the formation of these possible molecular candidates composed of two B mesons. However, such low-energy hadron interaction lie in a region which is very far away from the perturbative regime, precluding us to use the perturbative approach with the fundamental QCD Lagrangian. Therefore, we need some non-perturbative approaches, such as QCD sum rules(QCDSR)~\cite{Shifman,RRY,Nielsen}, to calculate the form factors. Besides, QCDSR at finite temperature illustrates mass shifts and width broadening~\cite{Su1,Su2}. Thus, it is expected that the form factor of $B^{\ast}B^{\ast}\rho$ vertex may be sensitive to high temperatures and it is maybe another interesting task.

In Ref.~\cite{LZH1}, the coupling constants $g_{B^{\ast}B^{\ast}\rho}$ and $f_{B^{\ast}B^{\ast}\rho}$ are estimated using QCD light cone sum rule (LCSR) method. In this article, the form factors and the coupling constants of the $B^{\ast}B^{\ast}\rho$ vertex is calculated in the framework of the three-point QCDSR. We notice that in the case of $D^{\ast}D^{\ast}\rho$ vertex, the form factor and coupling constant have been studied with three-point QCDSR using the effective Lagrangian based on $SU(4)$ flavor symmetry~\cite{bcnn08}. Different from above situations, we consider the effective Lagrangian, which is constructed based on the chiral symmetry. Herein, we use the same technique developed in the previous work for the evaluation of the couplings in the vetices $D^{\ast} D \pi$~\cite{nnbcs00,nnb02}, $D D \rho$~\cite{bclnn01}, $D^{\ast} D^{\ast} \pi$~\cite{cdnn05}, $D^{\ast}D^{\ast}\rho$~\cite{bcnn08}, $D D \omega$~\cite{hmm07}, $D^{\ast}_s D K^{\ast}(892)$~\cite{Azizi10}, $D_s D K^{\ast}_{0}$~\cite{Azizi11} and $B^{\ast}_{s1}B^{\ast}K$~\cite{Cui}.

This paper is organized as follows. In Sec.~\ref{sec2}, we give the details of QCDSR for the $B^{\ast}B^{\ast}\rho$ vertex when both $B^{\ast}$ and $\rho$ mesons are off-shell. Sec.~\ref{sec3} is devoted to the numerical analysis and discussion. Additionally, the Appendix presents the formula of form factors.

\section{The sum rule for the $B^{\ast}B^{\ast}\rho$ vertex}\label{sec2}
In this section, we give QCDSR for the form factors of the $B^{\ast}B^{\ast}\rho$ vertex.
The three-point function associated with the $B^{\ast}B^{\ast}\rho$ vertex,
for an off-shell $B^{\ast}$ meson, is given by
\begin{equation}
\Gamma_{\mu\nu\alpha}^{\bar B^{*0}}(p,p^{\prime})=\int d^4x \, d^4y \;\;
e^{ip^{\prime}\cdot x} \, e^{-iq\cdot y}
\langle 0|T\{j_{\mu}^{\rho^-}(x) j_{\nu}^{\bar B^{*0}}(y)
 j_{\alpha}^{B^{{\ast}\dagger}}
(0)|0\rangle,
\label{correboff}
\end{equation}
where the interpolating currents are $j_{\mu}^{\rho^-}(x) = \bar u(x) \gamma_{\mu} d(x)$, $j_{\nu}^{\bar B^{*0}}(x) = \bar d(x)\gamma_{\nu}b(x)$, and $j_{\alpha}^{B^{\ast}}(x)= \bar u(x) \gamma_{\alpha}b(x)$.
The correlation function for an off-shell ${\rho}$ meson is
\begin{equation}
\Gamma_{\mu\nu\alpha}^{\rho}(p,p^{\prime})=\int d^4x \,
d^4y \;\; e^{ip^{\prime}\cdot x} \, e^{-iq\cdot y}\;
\langle 0|T\{j_{\mu}^{\bar B^{*0}}(x) j_{\nu}^{\rho^-}(y)
 j_{\alpha}^{B^{\ast} \dagger}(0)\}|0\rangle\,.
\label{correroff}
\end{equation}
In the expressions, $q=p'-p$ is the transferred momentum. There are fourteen independent Lorentz structures in the general expression for the vertices (\ref{correboff}) and (\ref{correroff}). We can write $\Gamma_{\mu\nu\alpha}$ in terms of the invariant amplitudes associated with each one of these tensor structures in the following form:
\begin{eqnarray}
\Gamma_{\mu\nu\alpha}^{B^{\ast}}(p,p^{\prime})&=&
    \Gamma_1(p^2 , {p^{\prime}}^2 , q^2) g_{\mu \nu} p_{\alpha}
  + \Gamma_2(p^2,{p^{\prime}}^2, q^2) g_{\mu \alpha} p_{\nu}
  + \Gamma_3(p^2,{p^{\prime}}^2 , q^2) g_{\nu \alpha} p_{\mu}\nonumber \\
&&
  + \Gamma_4(p^2,{p^{\prime}}^2 ,q^2) g_{\mu \nu} p^{\prime}_{\alpha} + \Gamma_5(p^2, {p^{\prime}}^2 ,q^2) g_{\mu \alpha} p^{\prime}_{\nu}
  + \Gamma_6(p^2,{p^{\prime}}^2 ,q^2) g_{\nu\alpha} p^{\prime}_{\mu}\nonumber \\
&&
  + \Gamma_7(p^2,{p^{\prime}}^2 ,q^2) p_{\mu} p_{\nu} p_{\alpha}
  + \Gamma_8(p^2,{p^{\prime}}^2 ,q^2) p^{\prime}_{\mu} p_{\nu} p_{\alpha} + \Gamma_9(p^2,{p^{\prime}}^2 ,q^2) p_{\mu} p^{\prime}_{\nu} p_{\alpha}\nonumber \\
&&
  + \Gamma_{10}(p^2,{p^{\prime}}^2 ,q^2) p_{\mu} p_{\nu} p^{\prime}_{\alpha}
  + \Gamma_{11}(p^2,{p^{\prime}}^2 ,q^2) p^{\prime}_{\mu} p^{\prime}_{\nu} p_{\alpha}
  + \Gamma_{12}(p^2,{p^{\prime}}^2 ,q^2) p^{\prime}_{\mu} p_{\nu} p^{\prime}_{\alpha}  \nonumber \\
&&+ \Gamma_{13}(p^2,{p^{\prime}}^2 ,q^2) p_{\mu} p^{\prime}_{\nu}p^{\prime}_{\alpha}
  + \Gamma_{14}(p^2,{p^{\prime}}^2 ,q^2) p^{\prime}_{\mu} p^{\prime}_{\nu} p^{\prime}_{\alpha}.
\label{trace}
\end{eqnarray}

Due to $j_{\mu}^{\rho^-}$ is a conserved current, five constraints among these fourteen independent Lorentz structures are introduced. Therefore, only nine of them are independent. However, in the sense of calculating coupling constant, we can work with any one of the fourteen structures. There are some points that one must follow: (i) The chosen structure must appear in the phenomenological side. (ii) The chosen structure should exhibit good OPE (operator product expansion) convergence. (iii) The chosen structure should have a stability that guarantees a good match between the two sides of the sum rule. After the calculations, the structures that obey these points are $g_{\mu \nu} p^{\prime}_{\alpha}$ in the case $\rho$ off-shell and $g_{\mu \nu} p_{\alpha}$ in the case $B^{\ast}$ off-shell for $f_{B^{\ast}B^{\ast}\rho}$. Whereas, the structures are $g_{\alpha\nu} p^{\prime}_{\mu}$ in the case $B^{\ast}$ off-shell, and $g_{\alpha\mu} p^{\prime}_{\nu}$ in the case $\rho$ off-shell for $g_{B^{\ast}B^{\ast}\rho}$.

In order to get the sum rules, the correlation functions need to be calculated in two different ways: In phenomenological side, they are presented at the hadron level introducing hadronic parameters; in theoretical side,
they are calculated in terms of quark and gluon degrees of freedom by performing Wilson's OPE. The sum rules for the form factors are obtained with both representations being matched via quark-hadron duality and equating the coefficient of a sufficient structure from both sides of the same correlation functions. In order to improve the matching between the two side of the sum rules, double Borel transformation with respect to the variables, $P^2=-p^2\rightarrow M^2$ and ${P^\prime}^2=-{p^\prime}^2\rightarrow {M^{\prime}}^2$, is performed.

The phenomenological part of the first
correlation function (\ref{correboff}) is obtained by saturating the complete set of appropriate $B^{\ast}$
and $\rho$ states. The matrix elements associated with the
$B^{\ast}B^{\ast}\rho$ momentum dependent vertices can be deduced from Eq. (\ref{1}), which can be written in terms of the form factors:
\begin{eqnarray}
\langle \bar B^{{\ast}0}(p,\eta)\rho^-(q,\epsilon)|B^{{\ast}-}(p+q,\xi)\rangle &=& -2\sqrt{2}g_{B^{\ast}B^{\ast}\rho}(q^2)\left(\eta^{\ast}\cdot\xi\right)
\left(p\cdot\epsilon^{\ast}\right)\nonumber\\&&
-4\sqrt{2}f_{B^{\ast}B^{\ast}\rho}(q^2)m_{B^{\ast}}\left[\left(\eta^{\ast}\cdot\epsilon^{\ast}\right)\left(\xi\cdot
q\right)-\left(\xi\cdot\epsilon^{\ast}\right)\left(\eta^{\ast}\cdot
q\right)\right]\nonumber\\.
\label{ffb}
\end{eqnarray}
The meson decay constants $f_{B^{\ast}}$ and $f_{\rho}$ are
defined by the following matrix elements:
\begin{eqnarray}
\langle 0|j_{\nu}^{B^{\ast}}|{B^{\ast}(p)}\rangle&=& m_{B^{\ast}} f_{B^{\ast}}\epsilon^{\nu}_{B^{\ast}}(p),\nonumber\\
\langle 0|j_{\mu}^{\rho}|{\rho(p)}\rangle&=& m_{\rho} f_{\rho} \epsilon^{\mu}_{\rho}(p).
\label{frho}
\end{eqnarray}
Saturating Eq.~(\ref{correboff}) with
$B^{\ast}$ and $\rho$ states, and using Eqs.~(\ref{ffb}) and (\ref{frho}), then summing over polarization vectors via
\begin{eqnarray}
\epsilon^{\mu}_{\rho}(p){\epsilon^{\nu}_{\rho}}^{*}(p)=-g_{\mu\nu}+\frac{p_{\nu}p_{\mu}}{m_{\rho}^2},\nonumber\\
\epsilon^{\mu}_{B^{*}}(p){\epsilon^{\nu}_{B^{*}}}^{*}(p)=-g_{\mu\nu}+\frac{p_{\mu}p_{\nu}}{m_{B^{\ast}}^2},
\label{polvec}
\end{eqnarray}
the physical side of the correlation function for $B^{\ast}$ off-shell is obtained as
\begin{eqnarray}
\Gamma_{\mu\nu\alpha}^{(B^{\ast})phen}(p,p^{\prime})&=&\frac{m_{\rho}f_{\rho}f_{B^{\ast}}^2m_{B^{\ast}}^2}
{(P^2+m^2_{B^*})(Q^2+m^2_{B^*})({P^\prime}^2 +m^2_{\rho})}\times\nonumber\\
&&\left[-2\sqrt{2} g^{B^{\ast}}_{B^{\ast} B^{\ast}\rho}(q^2)(-g_{\nu\beta}+\frac{q_{\nu} q_{\beta}}{ m^2_{B^*}})(-g_{\alpha\beta}+\frac{p_{\alpha}p_{\beta}}{ m^2_{B^*}})(-g_{\mu\gamma}+\frac{p^{\prime}_{\mu}p^{\prime}_{\gamma}}{ m^2_{\rho}})q_{\gamma}\right.\nonumber\\
&&-4\sqrt{2} f^{B^{\ast}}_{B^{\ast} B^{\ast}\rho}(q^2)m_{B^*}(-g_{\nu\beta}+\frac{q_{\nu} q_{\beta}}{ m^2_{B^*}})(-g_{\alpha\gamma}+\frac{p_{\alpha}p_{\gamma}}{ m^2_{B^*}})(-g_{\mu\beta}+\frac{p^{\prime}_{\mu}p^{\prime}_{\beta}}{ m^2_{\rho}})p^{\prime}_{\gamma}\nonumber\\
&&\left.+4\sqrt{2} f^{B^{\ast}}_{B^{\ast} B^{\ast}\rho}(q^2)m_{B^*}(-g_{\nu\gamma}+\frac{q_\nu q_{\gamma}}{ m^2_{B^*}})(-g_{\alpha\beta}+\frac{p_{\alpha}p_{\beta}}{ m^2_{B^*}})(-g_{\mu\beta}+\frac{p^{\prime}_{\mu}p^{\prime}_{\beta}}{ m^2_{\rho}})p^{\prime}_{\gamma}\right] \nonumber\\
&&+...\,.
\label{phenboff}
\end{eqnarray}

In a similar way, we obtain the final expression of the physical side of the correlation function for an off-shell $\rho$ meson as:
\begin{eqnarray}
\Gamma_{\mu\nu\alpha}^{(\rho)phen}(p,p^{\prime})&=&\frac{m_{\rho}f_{\rho}f_{B^{\ast}}^2m_{B^{\ast}}^2}
{(P^2+m^2_{B^*})(Q^2+m^2_{\rho})({P^\prime}^2 +m^2_{B^*})}\times\nonumber\\
&&\left[-2\sqrt{2} g^{\rho}_{B^{\ast} B^{\ast}\rho}(q^2)(-g_{\mu\beta}+\frac{p^{\prime}_{\mu} p^{\prime}_{\beta}}{ m^2_{B^*}})(-g_{\alpha\beta}+\frac{p_{\alpha}p_{\beta}}{ m^2_{B^*}})(-g_{\nu\gamma}+\frac{q_{\nu}q_{\gamma}}{ m^2_{\rho}})p^{\prime}_{\gamma}\right.\nonumber\\
&&-4\sqrt{2} f^{\rho}_{B^{\ast} B^{\ast}\rho}(q^2)m_{B^*}(-g_{\mu\beta}+\frac{p^{\prime}_{\mu} p^{\prime}_{\beta}}{ m^2_{B^*}})(-g_{\alpha\gamma}+\frac{p_{\alpha}p_{\gamma}}{ m^2_{B^*}})(-g_{\nu\beta}+\frac{q_{\nu}q_{\beta}}{ m^2_{\rho}})q_{\gamma}\nonumber\\
&&\left.+4\sqrt{2} f^{\rho}_{B^{\ast} B^{\ast}\rho}(q^2)m_{B^*}(-g_{\mu\gamma}+\frac{q_\mu q_{\gamma}}{ m^2_{B^*}})(-g_{\alpha\beta}+\frac{p_{\alpha}p_{\beta}}{ m^2_{B^*}})(-g_{\nu\beta}+\frac{q_{\nu}q_{\beta}}{ m^2_{\rho}})q_{\gamma}\right] \nonumber\\
&&+...\,,
\label{phenrhooff}
\end{eqnarray}
where ``...." represents the contribution of the higher states and
continuum.

\begin{figure}
\begin{minipage}{7cm}
\epsfxsize=10cm \centerline{\epsffile{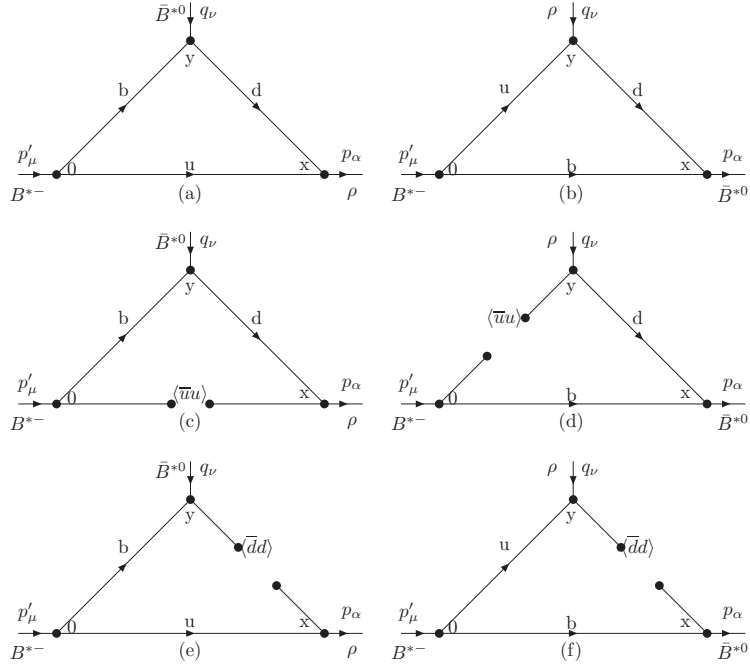}}
\end{minipage}
\caption{{(a) and (b): Bare loop diagram for the $B^{\ast}$  and
$\rho$ off-shell, respectively; (c) and (e): Diagrams
corresponding to quark condensate for the $B^{\ast}$ off-shell; (d) and
(f): Diagrams corresponding to quark condensate for the  $\rho$
off-shell.}}\label{Figure1}
\end{figure}

\begin{figure}
\begin{minipage}{7cm}
\epsfxsize=15cm \centerline{\epsffile{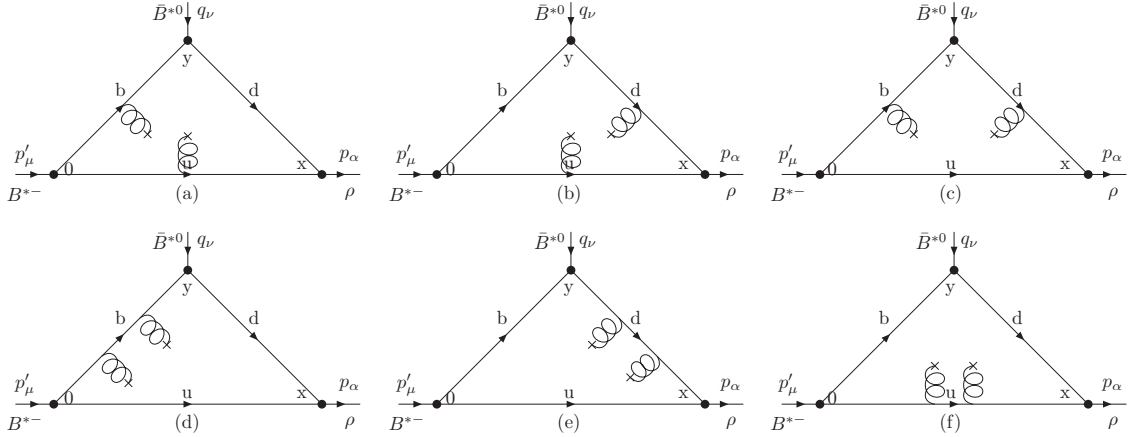}}
\end{minipage}
\caption{{Diagrams for contributions of bi-gluon operator in the case $B^{\ast}$ off-shell.}}\label{Figure2}
\end{figure}

\begin{figure}
\begin{minipage}{7cm}
\epsfxsize=15cm \centerline{\epsffile{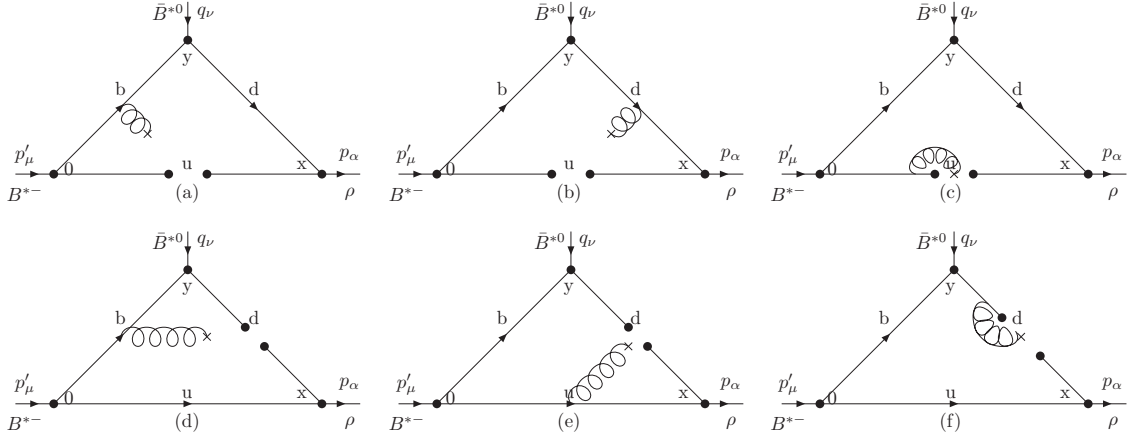}}
\end{minipage}
\caption{{Diagrams for mixed quark-gluon
operators in the case $B^{\ast}$ off-shell.}}\label{Figure3}
\end{figure}

In the following, we concentrate our attention to the QCD side of the correlation functions in the deep Euclidean space. The coefficients $\Gamma_{i}$ above can be written in terms of
perturbative and condensate terms
\begin{equation}
\Gamma_{i}=\Gamma_{i}^{per}+\Gamma_{i}^{(3)}+\Gamma_{i}^{(4)}+\Gamma_{i}^{(5)}
+\Gamma_{i}^{(6)}+\cdots
\label{gamma}
\end{equation}
where $\Gamma_{i}^{per}$ is the perturbative contribution, and $\Gamma_{i}^{(3)}$, $\cdots$, $\Gamma_{i}^{(5)}$ are
contributions of condensates of dimension 3, 4, 5, $\cdots$
operators in the OPE. The perturbative contribution and gluon
condensate contribution can be written in the form of dispersion
integration,
\begin{eqnarray}
\Gamma_{i}^{per}&=&-\frac{1}{4\pi^2}\int_{s_{min}}^\infty ds
\int_{u_{min}}^\infty du \:\frac{\rho^{per}_i(s,u,Q^2)}{(s-p^2)(u-{p^\prime}^2)}\;,
 \nonumber  \\
\Gamma_{i}^{(4)}&=&-\frac{1}{4\pi^2}\int_{s_{min}}^\infty ds
\int_{u_{min}}^\infty du \:\frac{\rho^{(4)}_{i}(s,u,Q^2)}{(s-p^2)(u-{p^\prime}^2)},
\nonumber  \nonumber\;\;\;\;\;\;i=1,\ldots,14, \label{dis}
\end{eqnarray}
where $\rho_i(s,u,Q^2)$ is the spectral density. The spectral density is obtained by calculating the bare
loop diagrams (a) and (b) in Fig.(\ref{Figure1}) for $B^{\ast}$ and
$\rho$ off-shell, respectively. In the calculation, Cutkosky rules are adopted to deal with the usual Feynman integral of these
diagrams, i.e., by replacing the quark propagators with Dirac
delta function $\frac{1}{q^2-m^2}\rightarrow (- 2\pi i)
\delta(q^2-m^2)\theta(q^0)$.
The integration region for the perturbative contribution
in Eq. (\ref{dis}) is determined from the facts that arguments of the
three $\delta$ functions must vanish simultaneously. The physical
regions of $s$ and $u$ are determined by the following
inequalities:\\
\begin{eqnarray}\label{13au}
-1\leq F^{B^{\ast}}(s,u)=\frac{2su+(s+u-t)(m_{b}^2-s)}
{\lambda^{1/2}(m_{b}^2,s,m_{u}^2)\lambda^{1/2}(s,u,t)}\leq+1,\nonumber\\
-1\leq F^{\rho}(s,u)=\frac{(s+u-t)(m_{b}^2+s)-2s(u+m_{b}^2)}
{|s-m_{b}^2|\lambda^{1/2}(s,u,t)}\leq+1,\nonumber\\
\end{eqnarray}
where $\lambda(a,b,c)=a^2+b^2+c^2-2ac-2bc-2ab$ and $t=q^2=-Q^2$.

The quark condensate contribution in the QCD side is determined from the quark condensate diagrams (c), (d), (e) and (f) of Fig.(\ref{Figure1}). As what has been shown in Refs.~\cite{nnbcs00,Nielsen}, heavy quark condensate contribution is negligible in comparison with the perturbative one. Thus, only light quark condensates contribute to the calculation. It is noticed that contributions of diagrams (d), (e) and (f) are zero after the double Borel transformation with respect to the both variables $P^2$ and ${P^\prime}^2$. Hence, we calculate the diagram (c) for the off-shell $B^{\ast}$ meson and obtain
\begin{eqnarray}\label{CorrelationFuncNonpert} \Pi_{\mu\nu\alpha}^{B^{\ast}(3)}&=&-\frac{m_{b}{\langle}\overline{u}u{\rangle}(-g_{\mu\nu}p'_{\alpha}+g_{\alpha\nu}p'_{\mu}+g_{\alpha\mu}p'_{\nu})}{(p^2-m_{b}^2)({p^{\prime}}^2)}.
\end{eqnarray}
The diagrams for the contribution of the gluon condensate in the case $B^{\ast}$ off-shell are depicted in Fig.(\ref{Figure2}). We follow the method employed in Refs.~\cite{yangmz,yangmz1}, namely, directly calculate the imaginary part of the integrals in terms of the Cutkosky¡¯s rule. The diagrams for the contribution of the quark-gluon mixing condensate in the case $B^{\ast}$ off-shell are depicted in Fig.(\ref{Figure3}). The results of the related Borel transformed coefficient $\hat{B}\Gamma_{i}$ in
Eq. (\ref{gamma}) are given in the appendix.

The quark-hadron duality assumption is adopted to subtract the contributions of the higher states and continuum, i.e., it is assumed that
\begin{eqnarray}\label{ope}
\rho^{higher states}(s,u) = \rho^{OPE}(s,u,t) \theta(s-s_0)
\theta(u-u_0),
\end{eqnarray}
where $s_0$ and $u_0$ are the continuum thresholds.

To improve the matching between the sides of the sum rules, the double Borel transformation are applied with respect to the $P^2=-p^2\rightarrow M^2$ and ${P^\prime}^2=-{p^\prime}^2\rightarrow {M^{\prime}}^2$. In this work we use the following relations between the Borel masses $M^2$ and ${M^{\prime}}^2$ which are $\frac{M^2}{{M^{\prime}}^2} = \frac{m^2_{\rho}}{m^2_{B^{\ast}}}$ for a  $B^{\ast}$ off-shell and $\frac{M^2}{M'^2} = 1$ for a $\rho$ off-shell.

\section{Numerical analysis}\label{sec3}
In the numerical analysis of the sum rules, input parameters are shown in Table~\ref{table1}. We first determine the three auxiliary parameters, namely the Borel mass parameter $M^2$ and the continuum thresholds, $s_0$ and $u_0$. The continuum thresholds, $s_0$ and $u_0$, are not completely arbitrary as they are correlated to the
energy of the first excited states with the same quantum numbers as the states we concern. They are given by $s_0=(m_{B^{\ast}} + \Delta_{s})^2$ and $u_0=(m+\Delta_{u})^2$, where $m$ is the $\rho$ meson mass for the case that $B^{\ast}$ is off-shell and the $B^{\ast}$ meson mass for that $\rho$ is off-shell. $ \Delta_u $ and $ \Delta_s$ are usually around $0.5 \; \mbox{GeV}$. The threshold $s_{0}$, $u_{0}$ and Borel parameter $M^{2}$ are varied to find the optimal stability window where pole dominance and OPE convergence of the sum rule are satisfied.
\begin{table}[h]
\caption{Parameters used in the calculation.}
\begin{center}
\begin{tabular}{ccccccccc}
\hline
$m_{b} (\mbox{GeV})$ &  $m_{B^{\ast}} (\mbox{GeV})$ & $m_{\rho} (\mbox{GeV})$ &  $f_{B^{\ast}}(\mbox{GeV})$  & $f_{\rho} (\mbox{GeV})$ & $\langle \bar u u \rangle(\mbox{GeV})^3$\\
\hline
$4.7 \pm 0.1$ &5.325 &0.775 &$0.16 \pm 0.01$ &$0.16 \pm 0.005$ &$(-0.23)^3$ \\ \hline
\end{tabular}
\end{center}\label{table1}
\end{table}
\subsection{QCD sum rules for $g_{B^{\ast}B^{\ast}\rho}(Q^2)$}
\begin{figure}
\begin{center}
\begin{minipage}[c]{0.5\textwidth}
\centering
\includegraphics[width=\textwidth]{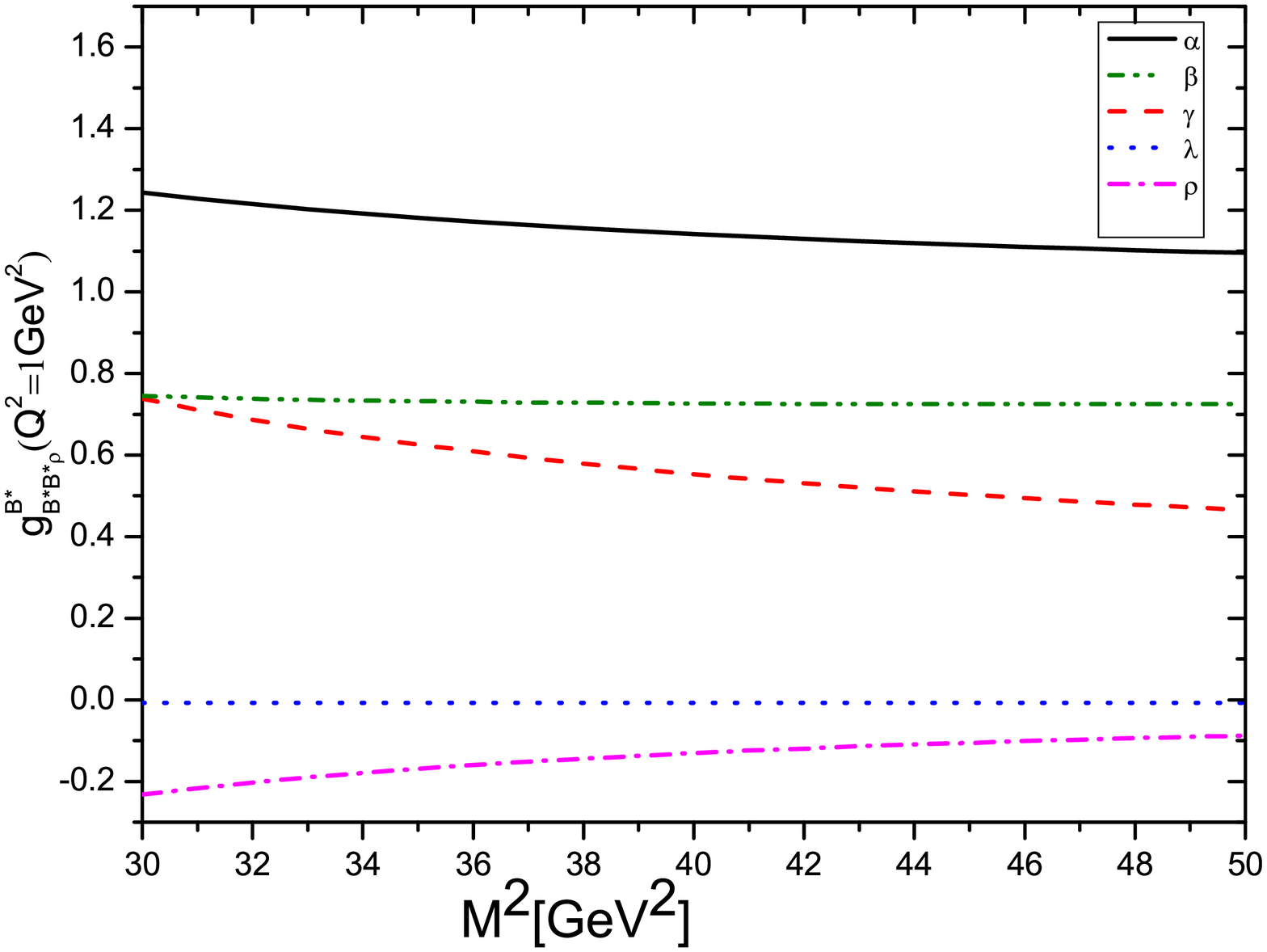}\\
(a)
\end{minipage}
\hspace{-0.1\textwidth}
\begin{minipage}[c]{0.5\textwidth}
\centering
\includegraphics[width=\textwidth]{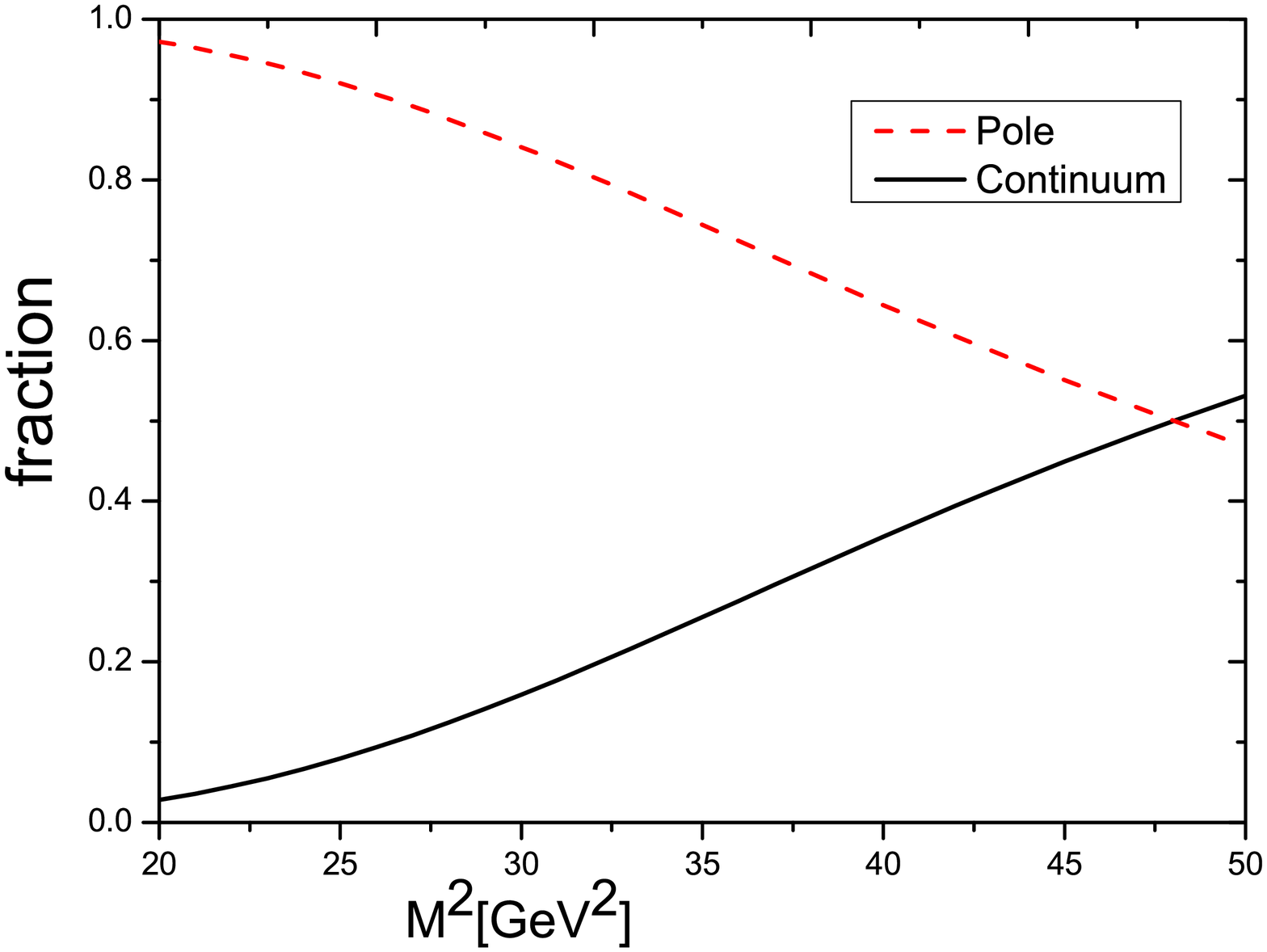}\\
(b)
\end{minipage}
\end{center}
\caption{a) The OPE convergence of the form factor $g^{B^{\ast}}_{B^{\ast}B^{\ast}\rho}(Q^2=1.0\,\mbox{GeV}^2)$ on Borel mass parameters $M^2$ for $\Delta_{s}=\Delta_{u} = 0.5\,\mbox{GeV}$. The notations $\alpha$, $\beta$, $\gamma$, $\lambda$ and $\rho$ correspond to total, perturbative, quark condensate, four-quark condensate and mixed condensate contribution respectively and b) pole-continuum contributions.}
\label{Figure4}
\end{figure}

Using $\Delta_{s}=\Delta_{u}=0.5\,\mbox{GeV} $ for the continuum thresholds and fixing $Q^2=1\,\mbox{GeV}^2$, we show the different contributions to the form factor $g^{B^{\ast}}_{B^{\ast}B^{\ast}\rho}$ as a function of the Borel variable, as can be seen in Fig.~(\ref{Figure4}a). We find an good OPE convergence and a good stability of $g^{B^{\ast}}_{B^{\ast}B^{\ast}\rho}$ for $M^2\geq 35\,\mbox{GeV}^2$. Fig.~(\ref{Figure4}b) demonstrates the contributions from the pole term and continuum term with variation of the Borel parameter $M^2$. We see that the pole contribution is larger than continuum contribution for $M^2\leq 47\,\mbox{GeV}^2$. We choose $M^2= 40\,\mbox{GeV}^2$ as a reference point.

\begin{figure}
\begin{minipage}{7cm}
\epsfxsize=10cm \centerline{\epsffile{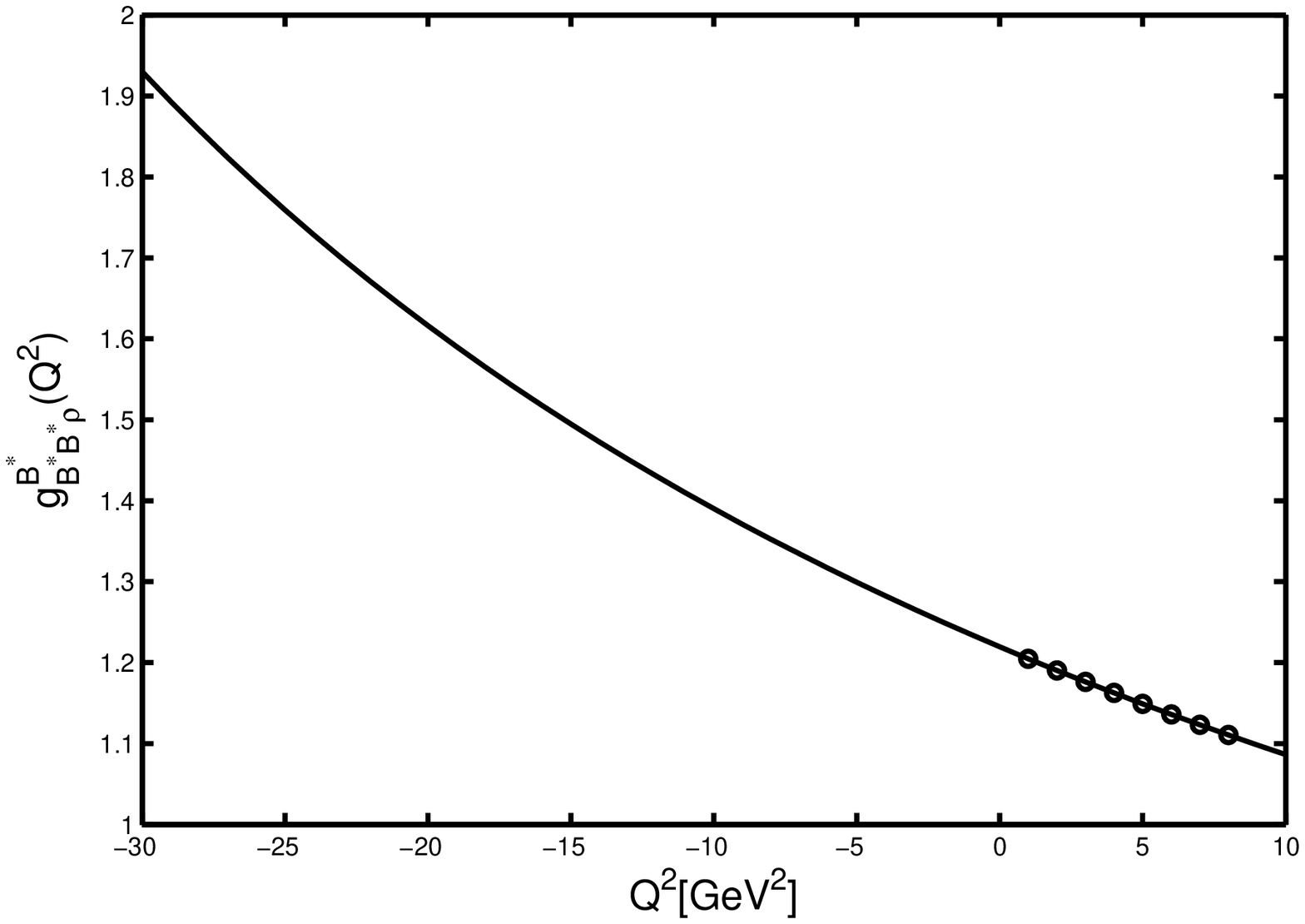}}
\end{minipage}
\caption{\quad $g^{B^{\ast}}_{B^{\ast}B^{\ast}\rho}(Q^2)$ (circles) QCDSR form factors as a function of
$Q^2$. The solid line correspond to the monopolar parametrization of the QCDSR data.}\label{Figure5}
\end{figure}

\begin{figure}
\begin{center}
\begin{minipage}[c]{0.5\textwidth}
\centering
\includegraphics[width=\textwidth]{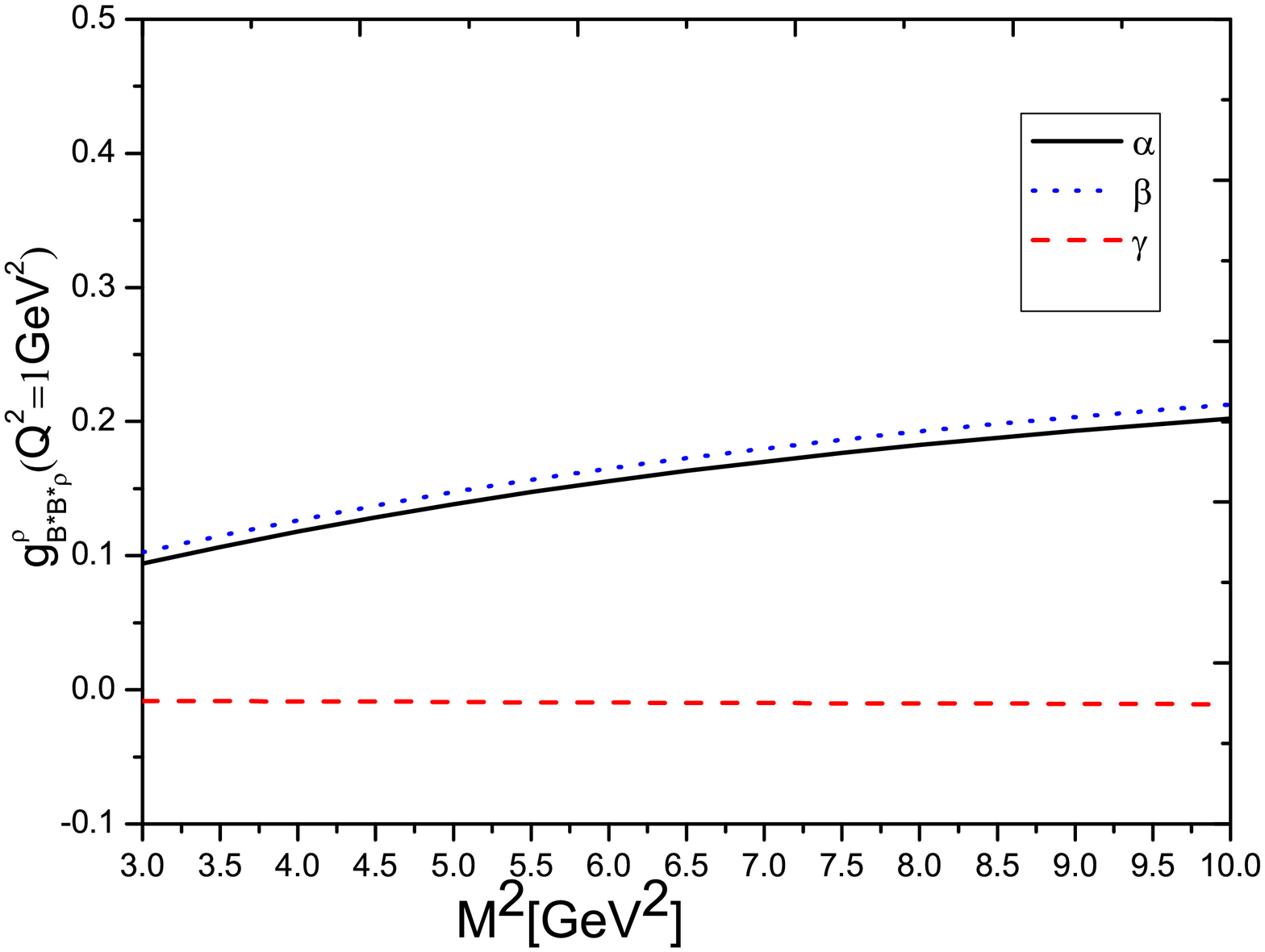}\\
(a)
\end{minipage}
\hspace{-0.1\textwidth}
\begin{minipage}[c]{0.5\textwidth}
\centering
\includegraphics[width=\textwidth]{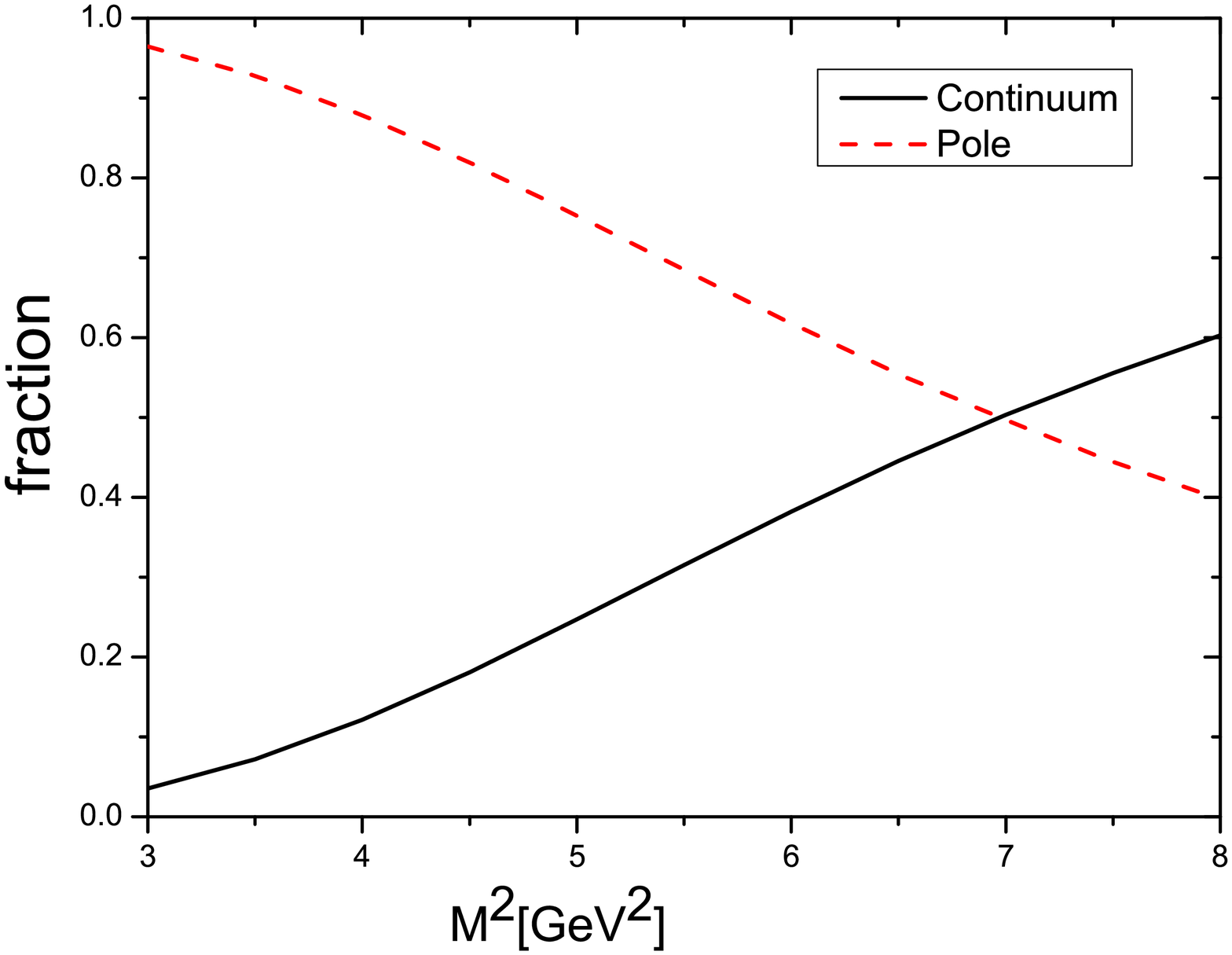}\\
(b)
\end{minipage}
\end{center}
\caption{a) The OPE convergence of the form factor $g^{\rho}_{B^{\ast}B^{\ast}\rho}(Q^2=1.0\,\mbox{GeV}^2)$ on Borel mass parameters $M^2$ for $\Delta_{s}=\Delta_{u} = 0.5\,\mbox{GeV}$. The notations $\alpha$, $\beta$ and $\gamma$ correspond to total, perturbative and four-quark condensate contributions respectively and b) pole-continuum contributions.}\label{Figure6}
\end{figure}

\begin{figure}
\begin{minipage}{7cm}
\epsfxsize=10cm \centerline{\epsffile{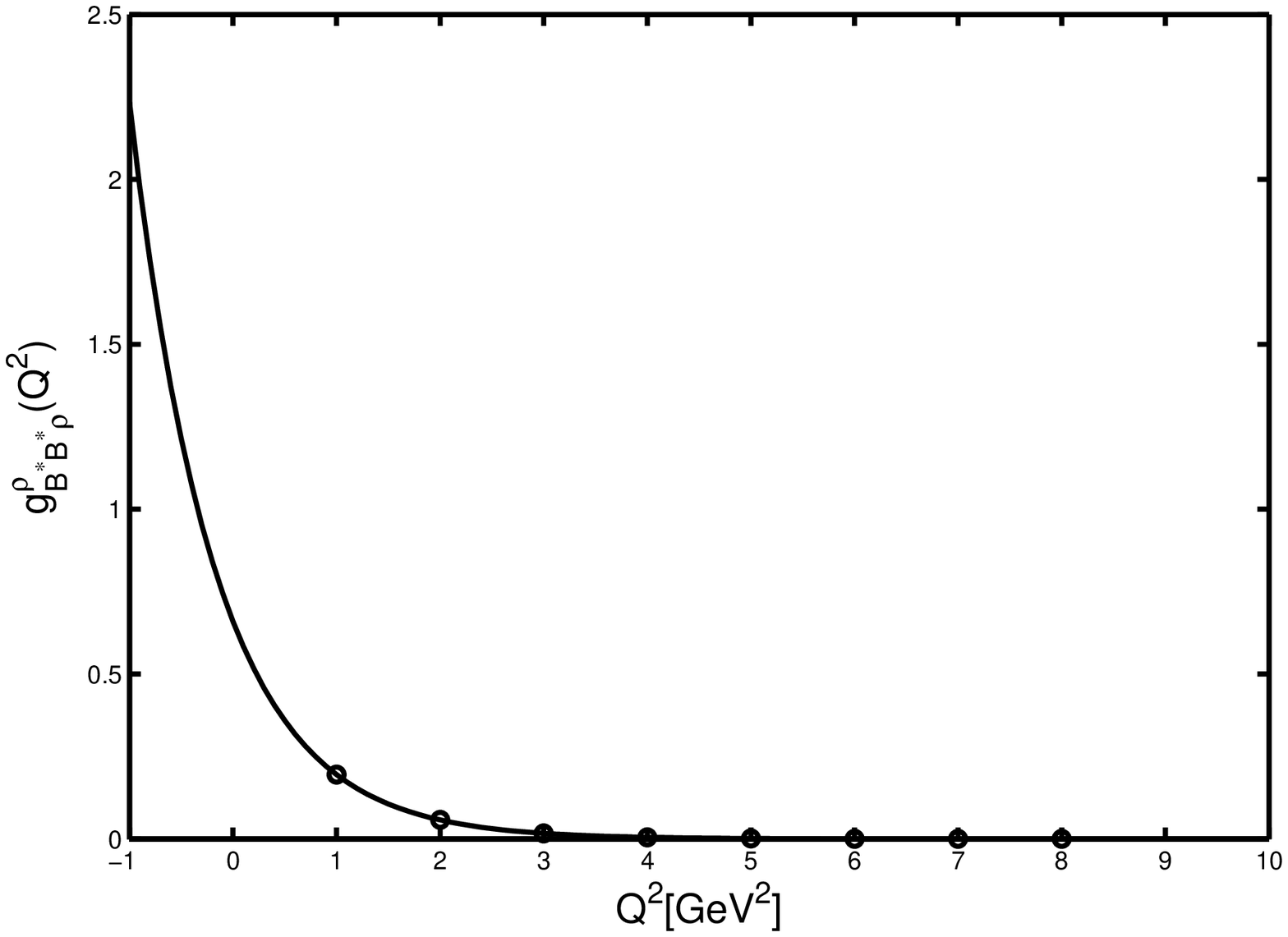}}
\end{minipage}
\caption{\quad $g^{\rho}_{B^{\ast}B^{\ast}\rho}(Q^2)$ (circles) QCDSR form factors as a function of
$Q^2$. The solid line correspond to the exponential parametrization of the QCDSR data.}\label{Figure7}
\end{figure}

Now, we would like to discuss the behavior of the form factors in terms of $Q^2$, which is plotted in Fig.~(\ref{Figure5}). In this figure, the circles correspond to the form factor $g^{B^{\ast}}_{B^{\ast}B^{\ast}\rho}(Q^2)$ in the interval where the sum rule is valid. Our result is better extrapolated by the mono-polar parametrization:
\begin{equation}
g_{B^{\ast}B^{\ast}\rho}^{B^{\ast}}(Q^2)= \frac{99.4\,\mbox{GeV}^2} {Q^2 + 75.5\,\mbox{GeV}^2} \;.
\label{gmonoboff}
\end{equation}
Coupling constant is defined as the value of the form factor at $Q^2=-m^2$, where $m$ is the mass of the off-shell meson. Using $Q^2=-m_{B^{\ast}}^2$ in Eq. (\ref{gmonoboff}), the coupling constant is obtained as $g^{B^{\ast}}_{B^{\ast}B^{\ast}\rho}=2.09$.

In the case $\rho$ off-shell, Fig.~(\ref{Figure6}a) demonstrates a good stability and OPE convergence of $g^{\rho}_{B^{\ast}B^{\ast}\rho}(Q^2=1.0\,\mbox{GeV}^2)$ with respect to the variations of Borel mass parameters for $M^2\geq 4\,\mbox{GeV}^2$. We see that the pole contribution is bigger than the continuum one in the Borel window $M^2\leq 7\,\mbox{GeV}^2$ from Fig.~(\ref{Figure6}b). We choose $M^2= 6.5\,\mbox{GeV}^2$. Our numerical results can be fitted by the exponential parametrization
\begin{eqnarray}\label{grhooff}
g^{\rho}_{B^{\ast}B^{\ast}\rho}(Q^2)=0.66~
\mbox{Exp}[\frac{-Q^2}{0.82\,\mbox{GeV}^2}],
\end{eqnarray}
shown by the solid line in Fig.(\ref{Figure7}). Also, $g^{\rho}_{B^{\ast}B^{\ast}\rho}=1.37$ is obtained at $Q^2=-m_{\rho}^2$ in Eq. (\ref{grhooff}). Taking the average of the two results, we get
\begin{eqnarray}\label{CoupConstg}
g_{B^{\ast}B^{\ast}\rho}=1.73\pm0.25.
\end{eqnarray}

Following the procedure of error estimate in Refs.~\cite{Nielsen1,Nielsen2}, with all parameters kept fixed, except one which is changed according to its intrinsic error, we calculate a new coupling constant and its deviation. Then we obtain percentage deviation related with each parameter and how sensitive this value is with respect to each parameter. Table~\ref{table2} show the percentage deviation for the two cases.

\begin{table}[h!]
\begin{center}
\begin{tabular}{|c|c|c|c|}\hline
               &    \multicolumn{2}{|c|}  {\textbf{Deviation {\%}}} \\ \hline
\textbf{Parameters} &\textbf{ $B^{\ast}$ off-shell}& \textbf{$\rho$ off-shell} \\ \hline
$f_{B^*}=160 \pm 10 $ (MeV)    &      $14.1$    & 15.2     \\\hline
  $f_{\rho}=160 \pm 5  $ (MeV)  & $3.2$ & $7.3 $     \\\hline
  $m_b=4.70\pm 0.1$ (GeV) & $15.1$ & $ 27.8$                  \\\hline
  $M^2 \pm 10\%$ (GeV)& $1.8$ & $1.7$               \\\hline
  $\Delta s \pm 0.1$ e $\Delta u \pm 0.1 $(GeV)  & $18.4$ &$ 23.5$  \\\hline
\end{tabular}
\caption{Percentage deviation related with each parameter for $g_{B^{\ast}B^{\ast}\rho}$. }
\label{table2}
\end{center}
\end{table}

Considering the uncertainties presented in the tables, the coupling constant is:
\begin{eqnarray}\label{CoupConstg1}
g_{B^{\ast}B^{\ast}\rho}=1.79\pm0.59.
\end{eqnarray}
\subsection{QCD sum rules for $f_{B^{\ast}B^{\ast}\rho}(Q^2)$}
\begin{figure}
\begin{center}
\begin{minipage}[c]{0.5\textwidth}
\centering
\includegraphics[width=\textwidth]{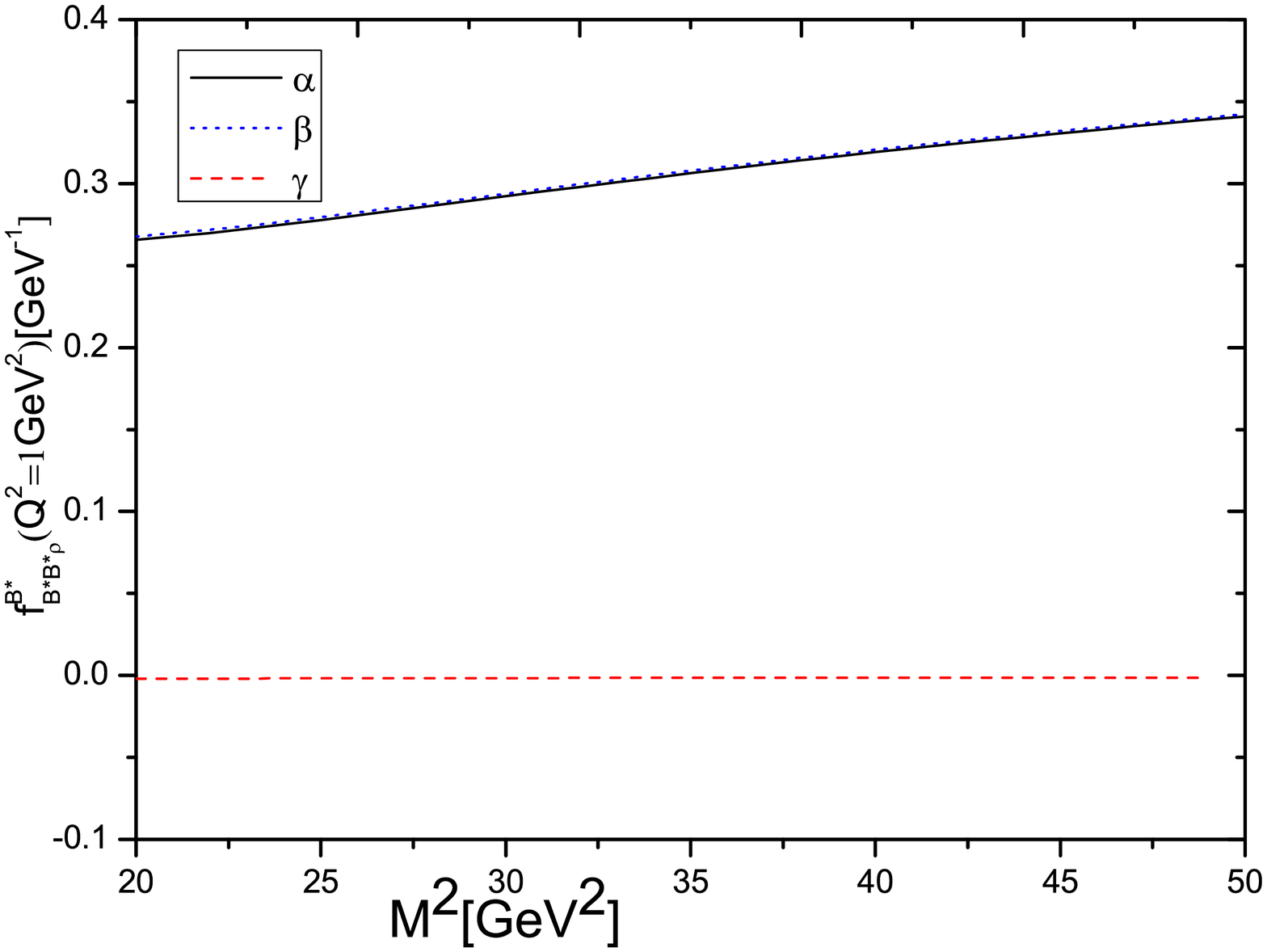}\\
(a)
\end{minipage}
\hspace{-0.1\textwidth}
\begin{minipage}[c]{0.5\textwidth}
\centering
\includegraphics[width=\textwidth]{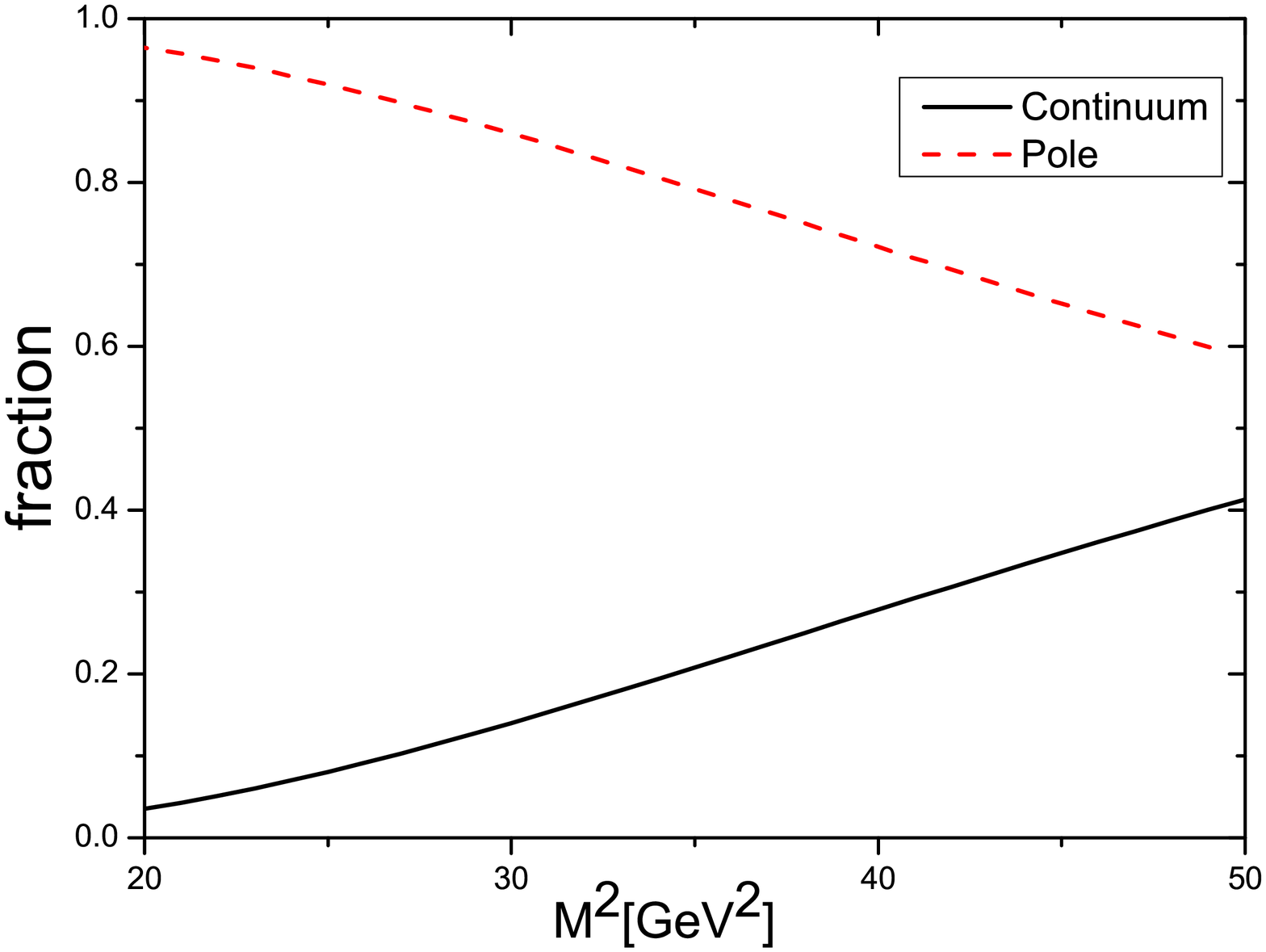}\\
(b)
\end{minipage}
\end{center}
\caption{a) The OPE convergence of the form factor $f^{B^{\ast}}_{B^{\ast}B^{\ast}\rho}(Q^2=1.0\,\mbox{GeV}^2)$ on Borel mass parameters $M^2$ for $\Delta_{s}=\Delta_{u} = 0.5\,\mbox{GeV}$. The notations $\alpha$, $\beta$ and $\gamma$ correspond to total, perturbative and four-quark condensate contributions respectively and b) pole-continuum contributions.}\label{Figure8}
\end{figure}

\begin{figure}
\begin{minipage}{7cm}
\epsfxsize=10cm \centerline{\epsffile{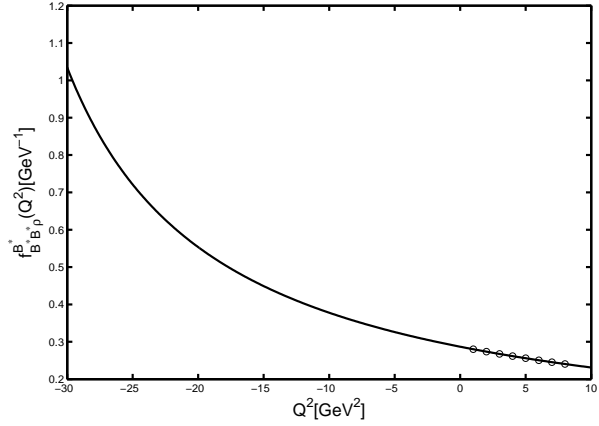}}
\end{minipage}
\caption{\quad $f^{B^{\ast}}_{B^{\ast}B^{\ast}\rho}(Q^2)$ (circles) QCDSR form factors as a function of
$Q^2$. The solid line correspond to the monopolar parametrization of the QCDSR data.}\label{Figure9}
\end{figure}

For $f^{B^{\ast}}_{B^{\ast}B^{\ast}\rho}(Q^2=1.0\,\mbox{GeV}^2)$, OPE convergence of the sum rule with the Borel mass and pole dominance are shown in Fig.(\ref{Figure8}). As the same procedure in the last subsection, the Borel mass is determined to be $ 25 \,\mbox{GeV}^2 $. With the thresholds $\Delta_s= 0.5 \,\mbox{GeV}$ and $\Delta_u = 0.5 \,\mbox{GeV} $, our numerical calculations of $f^{B^{\ast}}_{B^{\ast}B^{\ast}\rho}(Q^2)$ can be well fitted by the mono-polar parametrization shown in Fig.~(\ref{Figure9}):
\begin{equation}
f^{B^{\ast}}_{B^{\ast}B^{\ast}\rho}(Q^2)= \frac{13.5\,\mbox{GeV}} {Q^2 + 41.6\,\mbox{GeV}^2} \;.
\label{fmonoboff}
\end{equation}
Setting $Q^2=-m_{B^{\ast}}^2$ in Eq.~(\ref{fmonoboff}), the coupling constant is obtained as $f^{B^{\ast}}_{B^{\ast}B^{\ast}\rho}=1.01\,\mbox{GeV}^{-1}$.

\begin{figure}
\begin{center}
\begin{minipage}[c]{0.5\textwidth}
\centering
\includegraphics[width=\textwidth]{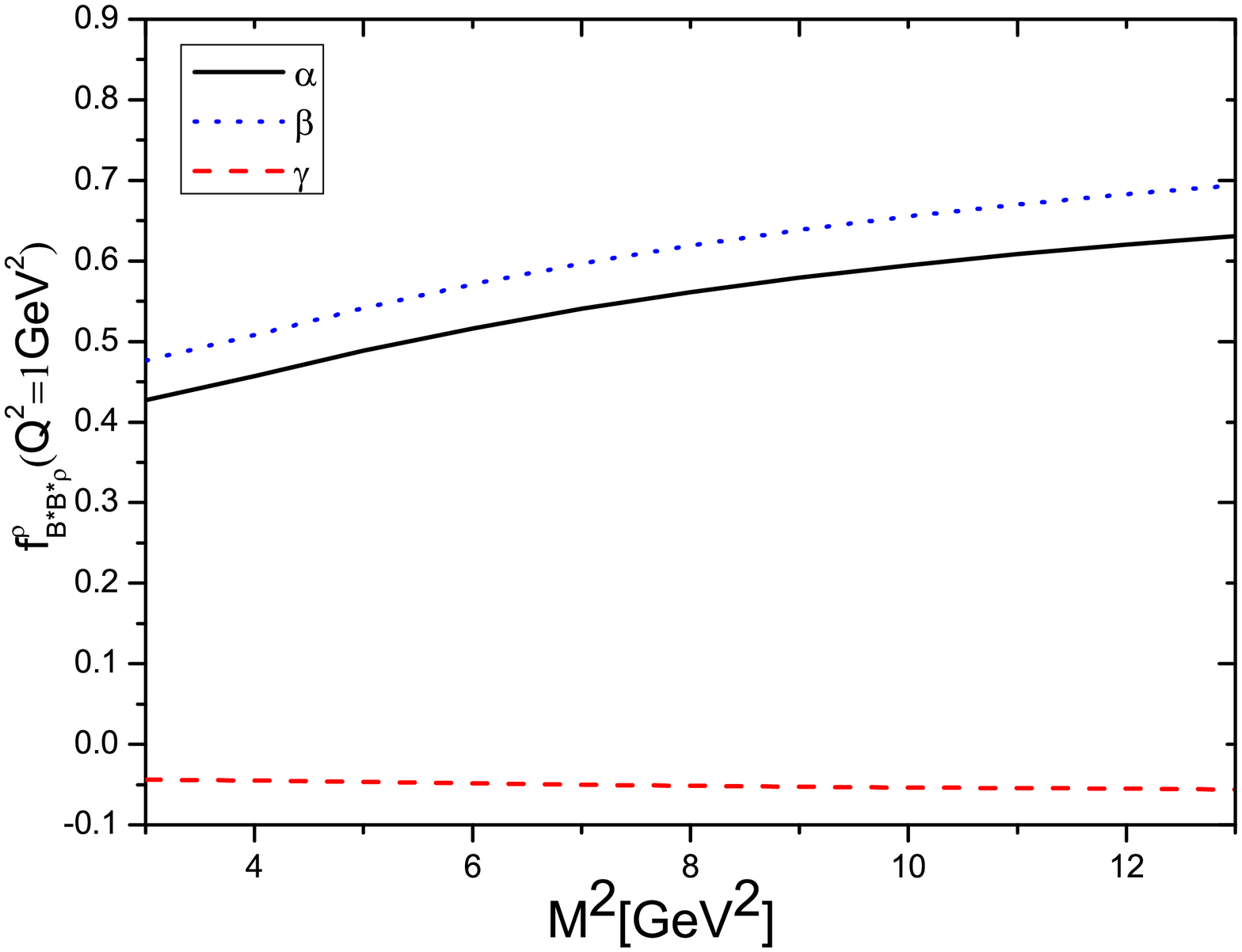}\\
(a)
\end{minipage}
\hspace{-0.1\textwidth}
\begin{minipage}[c]{0.5\textwidth}
\centering
\includegraphics[width=\textwidth]{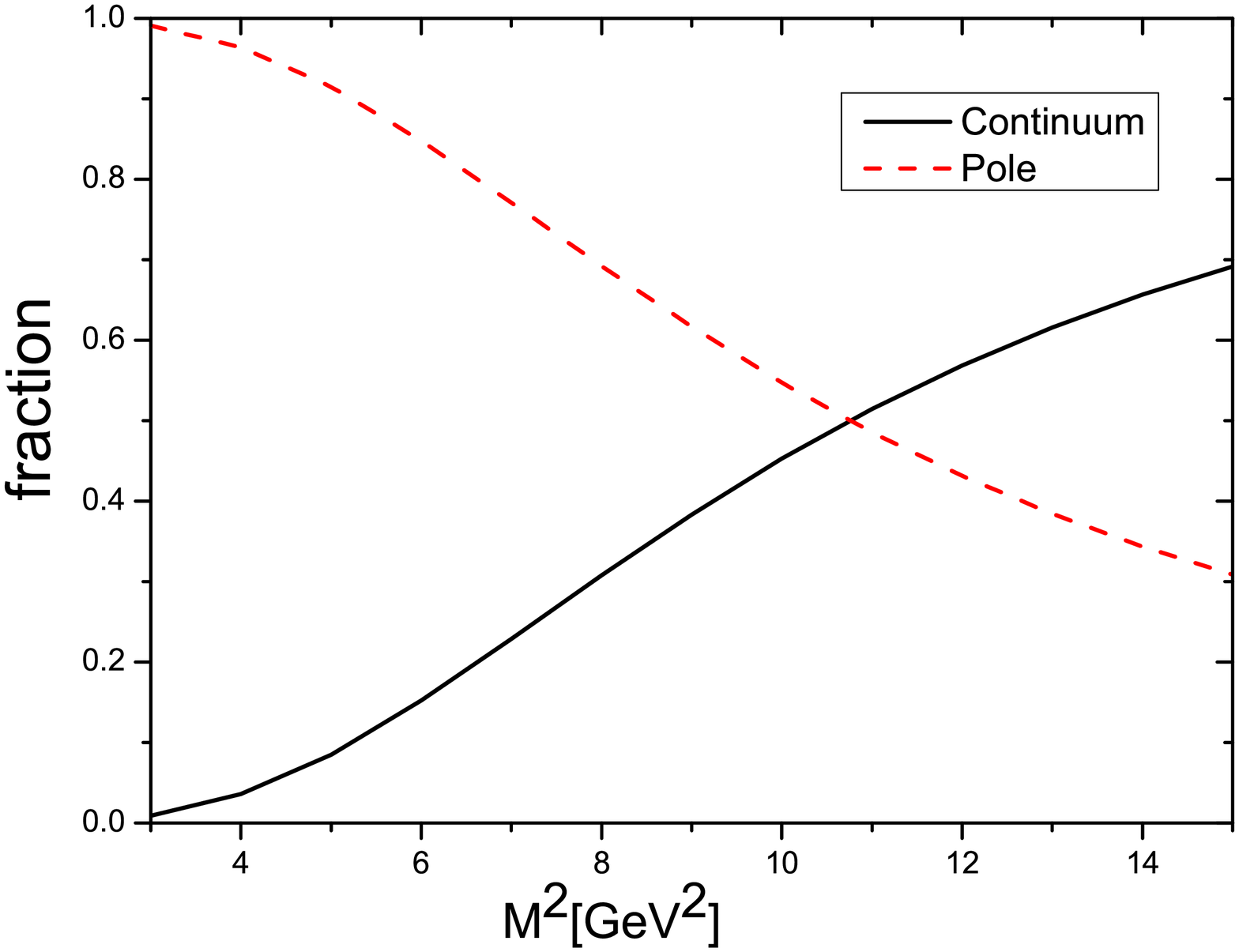}\\
(b)
\end{minipage}
\end{center}
\caption{a) The OPE convergence of the form factor $f^{\rho}_{B^{\ast}B^{\ast}\rho}(Q^2=1.0\,\mbox{GeV}^2)$ on Borel mass parameters $M^2$ for $\Delta_{s}=\Delta_{u} = 0.5\,\mbox{GeV}$. The notations $\alpha$, $\beta$ and $\gamma$ correspond to total, perturbative and four-quark condensate contributions respectively and b) pole-continuum contributions.}
\label{Figure10}
\end{figure}

\begin{figure}
\begin{minipage}{7cm}
\epsfxsize=10cm \centerline{\epsffile{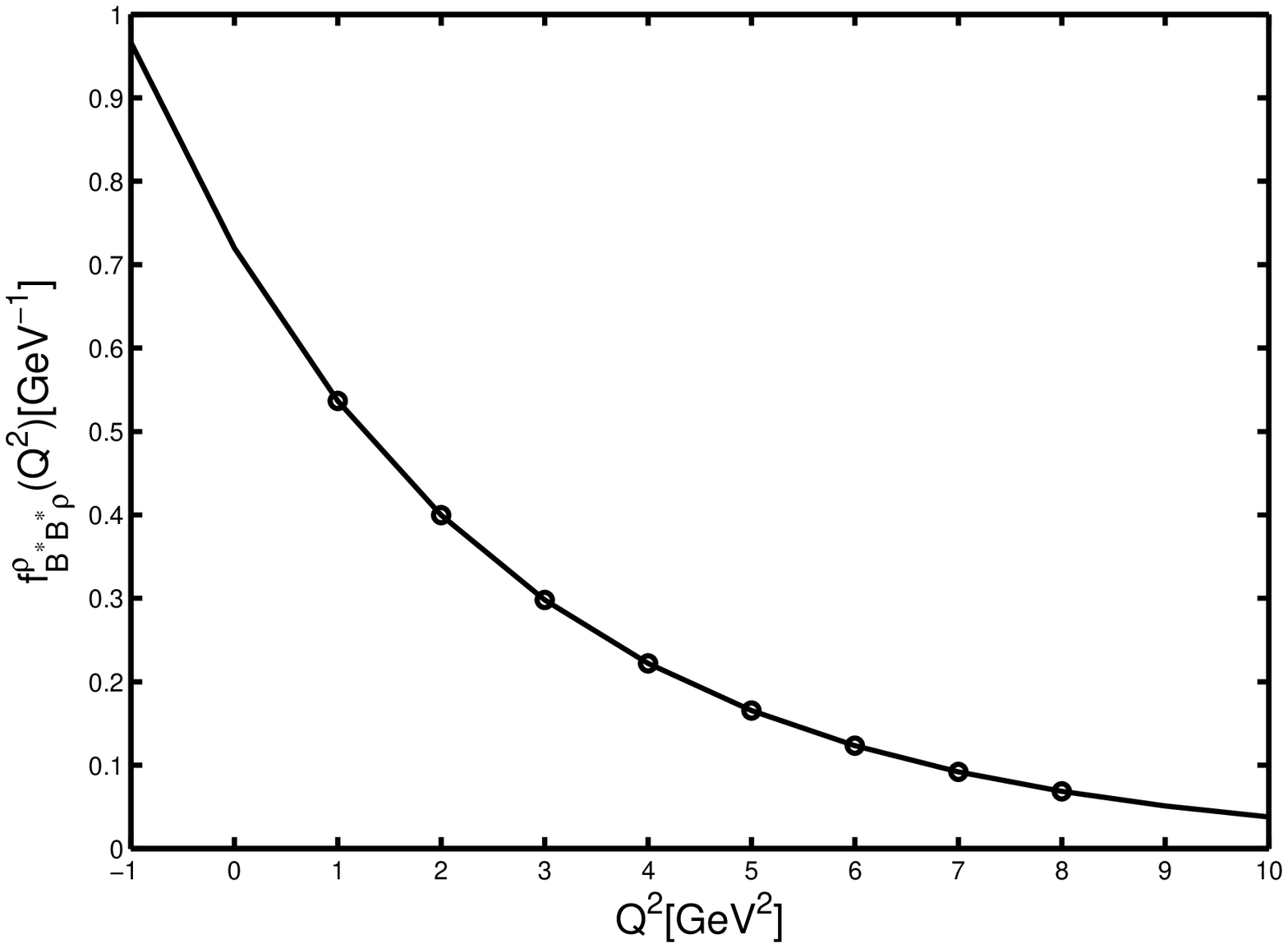}}
\end{minipage}
\caption{\quad $f^{\rho}_{B^{\ast}B^{\ast}\rho}(Q^2)$ (circles) QCDSR form factors as a function of
$Q^2$. The solid line correspond to the exponential parametrization of the QCDSR data.}
\label{Figure11}
\end{figure}

Figure~(\ref{Figure10}) is plotted to show the OPE convergence, stability and pole dominance of the sum rule with the Borel mass for $f^{\rho}_{B^{\ast}B^{\ast}\rho}(Q^2=1.0\,\mbox{GeV}^2)$. With the thresholds $\Delta_s= 0.5 \,\mbox{GeV}$ and $\Delta_u = 0.5 \,\mbox{GeV} $, and a Borel mass of $9\,\mbox{GeV}^2 $, our numerical calculations of $f^{\rho}_{B^{\ast}B^{\ast}\rho}(Q^2)$ can be well fitted by the exponential parametrization in Fig.~(\ref{Figure11}):
\begin{eqnarray}\label{frhooff}
f^{\rho}_{B^{\ast}B^{\ast}\rho}(Q^2)=0.72~
\mbox{Exp}\big[\frac{-Q^2}{3.44\mbox{GeV}^2}\big]~\mbox{GeV}^{-1},
\end{eqnarray}
Using $Q^2=-m_{\rho}^{2}$ in Eq.~(\ref{frhooff}),
the coupling constant is obtained as $f^{\rho}_{B^{\ast}B^{\ast}\rho}=0.86\,\mbox{GeV}^{-1}$. Taking the average of the two results, we get
\begin{eqnarray}\label{CoupConstg}
f_{B^{\ast}B^{\ast}\rho}=(0.94\pm0.08)\,\mbox{GeV}^{-1}.
\end{eqnarray}

Making the same procedure of error estimate as last subsection, Table~\ref{table3} show the percentage deviation for the coupling constant $f_{B^{\ast}B^{\ast}\rho}$.

\begin{table}[h!]
\begin{center}
\begin{tabular}{|c|c|c|c|}\hline
               &    \multicolumn{2}{|c|} {\textbf{Deviation {\%}}} \\\hline
\textbf{Parameters} &\textbf{ $B^{\ast}$ off-shell}& \textbf{$\rho$ off-shell} \\ \hline
$f_{B^*}=160 \pm 10 $ (MeV)    &      $15.5$    & 12.4     \\\hline
  $f_{\rho}=160 \pm 5  $ (MeV)  & $4.5$ & $3.7 $     \\\hline
  $m_b=4.70\pm 0.1$ (GeV) & $9.2$ & $ 23.4$                  \\\hline
  $M^2 \pm 10\%$ (GeV)& $2.8$ & $1.5$               \\\hline
  $\Delta s \pm 0.1$ e $\Delta u \pm 0.1 $(GeV)  & $14.4$ &$ 17.9$  \\\hline
\end{tabular}
\caption{Percentage deviation related with each parameter for $f_{B^{\ast}B^{\ast}\rho}$. }
\label{table3}
\end{center}
\end{table}

Considering the uncertainties presented in the tables, the coupling constant is:
\begin{eqnarray}\label{CoupConstg1}
f_{B^{\ast}B^{\ast}\rho}=(0.94\pm0.24)~\mbox{GeV}^{-1}.
\end{eqnarray}

\begin{table}
\caption{ Theoretical estimations of the strong coupling constants
from different models}
\begin{center}
\begin{tabular}{cccccccc}
\hline\hline
& Coupling constant & This work  &  \cite{LZH1}  & \\
\hline
&$g_{B^{\ast}B^{\ast}\rho}$  & $1.79\pm0.59$  & 1.88 & \\ \hline
&$f_{B^{\ast}B^{\ast}\rho}(\mbox{GeV}^{-1})$  & $0.94\pm0.24$  & 0.82 &\\ \hline
\hline
\end{tabular}
\end{center}
\label{table4}
\end{table}

It is noticed that the form factors obtained are different if the $B^{\ast}$ or the $\rho$ meson is off-shell but both give the compatible coupling constant. As commented in Ref.{\cite{Nielsen}}, the two sets of points ($B^{\ast}$ or $\rho$ off shell) can be fitted by different empirical formulas. However, the condition must be satisfied that when extrapolated to $Q^2=-m^2$, where $m$ is the mass of the off-shell meson, each fit should go to the compatible value of the coupling constant. Together with the predictions from LCSR~\cite{LZH1}, the numerical results of the coupling constant are presented in Table \ref{table2}. Comparison shows that our result $g_{B^{\ast}B^{\ast}\rho}$ and $f_{B^{\ast}B^{\ast}\rho}$ are in good agreement with their estimate. However, it is noticed that the authors only give the central values of $g_{B^{\ast}B^{\ast}\rho}$ and $f_{B^{\ast}B^{\ast}\rho}$, our results are compatible with their estimates in case that the uncertainties are considered in their work.

In summary, the form factors $g_{B^{\ast}B^{\ast}\rho}(Q^2)$ and $f_{B^{\ast}B^{\ast}\rho}(Q^2)$ parameterizing the $B^{\ast}B^{\ast}\rho$ vertex have been calculated in the framework of three-point QCDSR. Both cases that $B^{\ast}$ is off-shell and $\rho$ is off-shell have been considered. As a side product of the form factors, the coupling constants $g_{B^{\ast}B^{\ast}\rho}$ and $f_{B^{\ast}B^{\ast}\rho}$ are estimated, which are compatible with the results from the LCSR method~\cite{LZH1}. Due to the potential ability in analyzing absorption cross sections of $\Upsilon$ in experiments, the related events are expected to be observed in the LHC in the near future.

\section*{Acknowledgement}
This work was supported in part by the National Natural Science
Foundation of China under Contract Nos.10975184 and 11047117, 11105222.
\begin{center}{\bf Appendix}\end{center}

The Appendix is devoted to analytical results of the form factors. Two cases are considered, one is for $g^{B^{\ast}}_{B^{\ast}B^{\ast}\rho}(Q^2)$ and $g^{\rho}_{B^{\ast}B^{\ast}\rho}(Q^2)$, the other for $f^{B^{\ast}}_{B^{\ast}B^{\ast}\rho}(Q^2)$ and $f^{\rho}_{B^{\ast}B^{\ast}\rho}(Q^2)$.

{\bf (1)} For $g^{B^{\ast}}_{B^{\ast}B^{\ast}\rho}(Q^2)$, the form factor is
\begin{equation}
g^{B^{\ast}}_{B^{\ast}B^{\ast}\rho}(Q^2)=\frac{-m_{\rho}(Q^2+m_{B^{\ast}}^2)}{f_{B^{\ast}}^2 f_{\rho}m_{B^{\ast}}^2\sqrt{2}(m_{B^{\ast}}^2+m_{\rho}^2+Q^2)}
\mbox{e}^{\frac{m_{B^{\ast}}^2}{M^2}}\mbox{e}^{\frac{m_{\rho}^2}{{M^{\prime}}^2}}\hat{B}\Gamma,
\end{equation}

where
\begin{equation}
\hat{B}\Gamma=\hat{B}\Gamma^{pert}+\hat{B}\Gamma^{(3)}+\hat{B}\Gamma^{(4)}
+\hat{B}\Gamma^{(5)}.
\end{equation}

The perturbative contribution is
\begin{eqnarray}
\hat{B}\Gamma^{pert}&=&\frac{-1}{4~\pi^2}\int^{s_0}_{m_{b}^2}
ds\int_0^{u_{0}} du
\rho_g^{B^{\ast}(pert)}(s,u,t)\theta[1-F^{B^{\ast}}(s,u)^2]
e^{\frac{-s}{M^2}}e^{\frac{-u}
{{M^{\prime}}^2}},
\end{eqnarray}
where
\begin{eqnarray}
\rho_g^{B^{\ast}(pert)}(s,u,t)&=&\frac{3}{[\lambda(s,u,t)]^{5/2}}(s-t+u)\nonumber\\
&& (m_{b}^4-m_{b}^2 (s+t-u)+s t) (u (2 m_{b}^2+s+t)-(s-t)^2).
\end{eqnarray}

The condensate contributions are
\begin{eqnarray}
\hat{B}\Gamma^{(3)}&=&m_{b}{\langle}\overline{u}u{\rangle}
e^{-\frac{m_{b}^2}{M^2}},
\end{eqnarray}
\begin{eqnarray}
\hat{B}\Gamma^{(4)}&=&\frac{-1}{4~\pi^2}\int^{s_0}_{m_{b}^2}
ds\int_0^{u_{0}} du
\rho_g^{B^{\ast}(4)}(s,u,t)\theta[1-F^{B^{\ast}}(s,u)^2]
e^{\frac{-s}{M^2}}e^{\frac{-u}
{{M^{\prime}}^2}},
\end{eqnarray}
in which
\begin{eqnarray}
\rho_g^{B^{\ast}(4)}(s,t,u)&=&\frac{\langle g^{2}G^{2}\rangle}{2[\lambda(s,u,t)]^{5/2}}
\nonumber\\&&((s^3+7s(t-u)^2-11s^2(t + u)+3(t-u)^2(t+u)
\nonumber\\&&-4m_{b}^2(-2s^2+s(t-2u)+(t-u)(t+2u))),
\end{eqnarray}
and
\begin{eqnarray}
\hat{B}\Gamma^{(5)}&=&\frac{m_{b}\langle g\bar{q}\sigma\cdot G q\rangle}{4{M^{\prime}}^2}
 e^{-\frac{m_{b}^2}{M^2}}.
\end{eqnarray}

For $g^{\rho}_{B^{\ast}B^{\ast}\rho}(Q^2)$, the form factor is

\begin{eqnarray}
g^{\rho}_{B^{\ast}B^{\ast}\rho}(Q^2)&=&\frac{m_{\rho}(Q^2+{m_{\rho}}^2)}
{f_{B^{\ast}}^2 f_{\rho}m_{B^{\ast}}^2(2\sqrt{2}m_{\rho}^2 + \sqrt{2}Q^2)}
\mbox{e}^{\frac{m_{B^{\ast}}^2}{M^2}}\mbox{e}^{\frac{m_{B^{\ast}}^2}{{M^{\prime}}^2}}
\hat{B}\Gamma,
\end{eqnarray}
where
\begin{equation}
\hat{B}\Gamma=\hat{B}\Gamma^{per}+\hat{B}\Gamma^{(3)}+\hat{B}\Gamma^{(4)}
+\hat{B}\Gamma^{(5)}.
\end{equation}

The perturbative contribution is
\begin{eqnarray}
\hat{B}\Gamma^{per}&=&\frac{-1}{4~\pi^2}\int^{s_0}_{m_{b}^2}
ds\int^{u_0}_{m_{b}^2} du
\rho_g^{\rho(per)}(s,u,t)\theta[1-F^{\rho}(s,u)^2]
e^{\frac{-s}{M^2}}e^{\frac{-u}
{{M^{\prime}}^2}},
\end{eqnarray}
where
\begin{eqnarray}
\rho_g^{\rho(per)}(s,u,t)&=&\frac{3}{[\lambda(s,u,t)]^{5/2}}(s+t-u)\nonumber\\
&&(m_{b}^4-m_{b}^2(s-t+u)+su)(-2m_{b}^2t+s^2-s(t+2 u)+u(u-t)).
\end{eqnarray}
The condensate contributions are
\begin{eqnarray}
\hat{B}\Gamma^{(3)}&=&0,
\end{eqnarray}
\begin{eqnarray}
\hat{B}\Gamma^{(4)}&=&\frac{-1}{4~\pi^2}\int^{s_0}_{m_{b}^2}
ds\int_0^{u_{0}} du
\rho_g^{\rho(4)}(s,u,t)\theta[1-F^{\rho}(s,u)^2]
e^{\frac{-s}{M^2}}e^{\frac{-u}
{{M^{\prime}}^2}},
\end{eqnarray}
in which
\begin{eqnarray}
\rho_g^{\rho(4)}(s,t,u)&=&\frac{\langle g^{2}G^{2}\rangle}{2[\lambda(s,u,t)]^{5/2}}\nonumber\\
&&(-7s^3+(t-u)^2(t+5u)+s^2(3t+7u)+s(3t^2+2tu-5u^2)\nonumber\\
&&+m_{b}^2(8s^2+8st+8t^2-4su-4tu-4u^2)),
\end{eqnarray}
and
\begin{eqnarray}
\hat{B}\Gamma^{(5)}&=&0.
\end{eqnarray}

{\bf (2)} For $f^{B^{\ast}}_{B^{\ast}B^{\ast}\rho}(Q^2)$, the form factor is

\begin{eqnarray}\label{CoupCons-FBBR-Roffshel}
f^{B^{\ast}}_{B^{\ast}B^{\ast}\rho}(Q^2)&=&\frac{-(Q^2+m_{B^{\ast}}^2)}{f_{B^{\ast}}^2 f_{\rho}m_{B^{\ast}}m_{\rho}\sqrt{2}(m_{B^{\ast}}^2+m_{\rho}^2+Q^2)}
\mbox{e}^{\frac{m_{B^{\ast}}^2}{M^2}}\mbox{e}^{\frac{m_{\rho}^2}{{M^{\prime}}^2}}\hat{B}\Gamma
\end{eqnarray}
where
\begin{equation}
\hat{B}\Gamma=\hat{B}\Gamma^{per}+\hat{B}\Gamma^{(3)}+\hat{B}\Gamma^{(4)}
+\hat{B}\Gamma^{(5)}.
\end{equation}

The perturbative contribution is
\begin{eqnarray}
\hat{B}\Gamma^{per}&=&\left[\frac{-1}{4~\pi^2}\int^{s_0}_{m_{b}^2}
ds\int_0^{u_{0}} du
\rho_f^{B^{\ast}(per)}(s,u,t) \right.
\nonumber \\
&& \left.\theta[1-F^{B^{\ast}}(s,u)^2]
e^{\frac{-s}{M^2}}e^{\frac{-u}
{{M^{\prime}}^2}}\right],
\end{eqnarray}
where
\begin{eqnarray}
\rho_f^{B^{\ast}(per)}(s,t,u)&=&\frac{3u}{[\lambda(s,u,t)]^{5/2}}(4 m_{b}^6 u+m_{b}^4 (-s^2+2 s t-4 s u-t^2-4 t u+5 u^2)\nonumber\\
&&+m_{b}^2(2 s^2 t+2 s^2 u-4 s t^2+4 s t u-4 s u^2+2 t^3-2 t^2 u-2 t u^2+2 u^3)\nonumber\\
&&-s^2 t^2-s^2 t u+2 s t^3-2 s t^2 u-t^4+3 t^3 u-3 t^2 u^2+t u^3 ).
\end{eqnarray}

The condensate contributions are
\begin{eqnarray}
\hat{B}\Gamma^{(3)}&=&0,
\end{eqnarray}
\begin{eqnarray}
\hat{B}\Gamma^{(4)}&=&\frac{-1}{4~\pi^2}\int^{s_0}_{m_{b}^2}
ds\int_0^{u_{0}} du
\rho_f^{B^{\ast}(4)}(s,u,t)\theta[1-F^{B^{\ast}}(s,u)^2]
e^{\frac{-s}{M^2}}e^{\frac{-u}
{{M^{\prime}}^2}},
\end{eqnarray}
in which
\begin{eqnarray}
\rho_f^{B^{\ast}(4)}(s,t,u)&=&\frac{\langle g^{2}G^{2}\rangle}{2[\lambda(s,u,t)]^{5/2}}\nonumber\\
&&(-3s^3-3t^3+s^2(3t-7u)-7t^2u+11tu^2-u^3+s(3t^2+14tu+11u^2)\nonumber\\
&&+4m_{b}^2(s^2+t^2+tu+4u^2+s(-2t+u))),
\end{eqnarray}
and
\begin{eqnarray}
\hat{B}\Gamma^{(5)}&=&0.
\end{eqnarray}

For $f^{\rho}_{B^{\ast}B^{\ast}\rho}(Q^2)$, the form factor is
\begin{equation}
f^{\rho}_{B^{\ast}B^{\ast}\rho}(Q^2)=\frac{-(Q^2+{m_{\rho}}^2)}
{4\sqrt{2}f_{B^{\ast}}^2 f_{\rho}m_{B^{\ast}}^3m_{\rho}}
\mbox{e}^{\frac{m_{B^{\ast}}^2}{M^2}}\mbox{e}^{\frac{m_{B^{\ast}}^2}{{M^{\prime}}^2}}\hat{B}\Gamma,
\end{equation}
where
\begin{equation}
\hat{B}\Gamma=\hat{B}\Gamma^{per}+\hat{B}\Gamma^{(3)}+\hat{B}\Gamma^{(4)}
+\hat{B}\Gamma^{(5)}.
\end{equation}

The perturbative contribution is
\begin{eqnarray}
\hat{B}\Gamma^{per}&=&\frac{-1}{4~\pi^2}\int^{s_0}_{m_{b}^2}
ds\int^{u_0}_{m_{b}^2} du
\rho_f^{\rho(per)}(s,u,t) \theta[1-F^{\rho}(s,u)^2]
e^{\frac{-s}{M^2}}e^{\frac{-u}
{{M^{\prime}}^2}},
\end{eqnarray}
where
\begin{eqnarray}
\rho_f^{\rho(per)}(s,t,u)&=&\frac{3}{[\lambda(s,u,t)]^{5/2}}
(t (-2 m_{b}^6 (s+t-u)+m_{b}^4 (4 s^2-2 s (t+u)-2 (t-u)^2)\nonumber \\
&&-m_{b}^2 (s+t-u) ((s-t)^2+4 s u-2 t u+u^2)+2 s u (s (t+u)-(t-u)^2))).\nonumber\\
\end{eqnarray}

The condensate contributions are
\begin{eqnarray}
\hat{B}\Gamma^{(3)}&=&0,
\end{eqnarray}
\begin{eqnarray}
\hat{B}\Gamma^{(4)}&=&\frac{-1}{4~\pi^2}\int^{s_0}_{m_{b}^2}
ds\int^{s_0}_{m_{b}^2} du
\rho_f^{\rho(4)}(s,u,t)\theta[1-F^{\rho}(s,u)^2]
e^{\frac{-s}{M^2}}e^{\frac{-u}
{{M^{\prime}}^2}},
\end{eqnarray}
in which
\begin{eqnarray}
\rho_f^{\rho(4)}(s,t,u)&=&\frac{\langle g^{2}G^{2}\rangle}{2[\lambda(s,u,t)]^{5/2}}\nonumber\\
&&(-9s^3+(t-u)^2(5t-u)+s^2(11t+5u)\nonumber\\
&&+M_b^2(8s^2+8st+8t^2-4su-4tu-4u^2)+s(-7t^2+10tu+5u^2)),
\end{eqnarray}
and
\begin{eqnarray}
\hat{B}\Gamma^{(5)}&=&0.
\end{eqnarray}

\end{document}